\documentclass[pra,aip,reprint]{revtex4-1}

\usepackage{graphicx,amsmath,braket}
\usepackage{appendix}

\begin{document}

\title{Phase-coherent sensing of the center-of-mass motion of trapped-ion crystals}

\author{M.~Affolter}
\affiliation{National Institute of Standards and Technology, Boulder, Colorado 80305, USA}
\author{K.~A.~Gilmore}
\affiliation{National Institute of Standards and Technology, Boulder, Colorado 80305, USA}
\affiliation{Department of Physics, University of Colorado, Boulder, Colorado 80309, USA}
\author{J.~E.~Jordan}
\affiliation{National Institute of Standards and Technology, Boulder, Colorado 80305, USA}
\affiliation{Department of Physics, University of Colorado, Boulder, Colorado 80309, USA}
\author{J.~J.~Bollinger}
\affiliation{National Institute of Standards and Technology, Boulder, Colorado 80305, USA}

\date{\today}

\begin{abstract}
Trapped ions are sensitive detectors of weak forces and electric fields that excite ion motion.  Here measurements of the center-of-mass motion of a trapped-ion crystal that are phase-coherent with an applied weak external force are reported.  These experiments are conducted far from the trap motional frequency on a two-dimensional trapped-ion crystal of approximately 100 ions, and determine the fundamental measurement imprecision of our protocol free from noise associated with the center-of-mass mode.  The driven sinusoidal displacement of the crystal is detected by coupling the ion crystal motion to the internal spin-degree of freedom of the ions using an oscillating spin-dependent optical dipole force.  The resulting induced spin-precession is proportional to the displacement amplitude of the crystal, and is measured with near-projection-noise-limited resolution.  A $49\,$pm displacement is detected with a single measurement signal-to-noise ratio of 1, which is an order-of-magnitude improvement over prior phase-incoherent experiments.  This displacement amplitude is $40$ times smaller than the zero-point fluctuations.  With our repetition rate, a $8.4\,$pm$/\sqrt{\mathrm{Hz}}$ displacement sensitivity is achieved, which implies $12\,$yN$/\mathrm{ion}/\sqrt{\mathrm{Hz}}$ and $77\,\mu$V$/$m$/\sqrt{\mathrm{Hz}}$ sensitivities to forces and electric fields, respectively. This displacement sensitivity, when applied on-resonance with the center-of-mass mode, indicates the possibility of weak force and electric field detection below $10^{-3}\,$yN/ion and $1\,$nV/m, respectively.
\end{abstract}

\pacs{}

\maketitle

\section{Introduction}
\label{Intro}

The past several decades have seen an immense improvement in the ability to detect the motion of a mechanical oscillator.  From large scale gravitational wave detectors~\cite{GravWave}, to mesoscopic optomechanical resonators~\cite{AspelmeyerOpto, WilsonOpto, KampelOpto, SchrepplerOpto, MasonOpto}, to the small scale single-ion sensors~\cite{MaiwaldIon, HempelIon, ShanivIon, BurdPara}, new avenues of fundamental and applied physics have been opened as the limits of amplitude sensing have improved.  Outside of the practical applications of better sensors, the limits to which a displacement smaller than the ground state wave function can be measured is of fundamental interest in quantum metrology.  Experiments are now routinely able to measure displacements smaller than the zero-point motion~\cite{MasonOpto, WilsonOpto, HempelIon, BurdPara, GilmoreSensing}.

Ions trapped in a harmonic potential are a natural platform to explore the fundamental limits of amplitude sensing.  These systems have tunable frequencies, high quality factors $Q\sim10^{6}$, and can be cooled to near their motional ground state via laser cooling.  Measurements of weakly driven coherent amplitudes, both smaller and larger than the zero-point fluctuations, have been demonstrated in traps with very small numbers of ions~\cite{HempelIon, ShanivIon, BurdPara, WolfFock}, but few sensing experiments have been conducted on larger trapped-ion crystals~\cite{GilmoreSensing, BiercukMultIon}.  Larger ion crystals have the benefit of reduced spin-projection noise improving the sensitivity for detecting weak forces and electric fields.

Here we discuss a technique for detecting center-of-mass (COM) displacements orders-of-magnitude smaller than the amplitude of ground-state zero-point fluctuations $Z_{ZPT}$ of a two-dimensional trapped-ion crystal of approximately $100$~ions.  By applying a weak sinusoidal electric field to the ions, a driven COM oscillation $Z_{c}\cos(\omega t)$ of the crystal occurs with frequency $\omega$ and displacement amplitude $Z_{c}$.  This motion is then detected by coupling the driven displacement to the internal spin degree of freedom of the ions through an oscillating spin-dependent force $F_{0}\cos(\mu t)$ and measuring the induced spin-precession, which is proportional to $Z_{c}$.

In a previous implementation of this basic protocol~\cite{GilmoreSensing}, the relative phase between the driven motion and spin-dependent force varied from one realization of the experiment to the next (shot-to-shot variations).  This work extends those results by phase stabilizing the spin-dependent force and driven motion.  We discuss how the relative phase stability is measured, and the experimental improvements undertaken to achieve shot-to-shot phase noise of less than 5 degrees.  This phase stability improves the displacement sensitivity by an order-of-magnitude, and will enable new avenues of research including parametric amplification of the spin-dependent force~\cite{BurdPara, GePara}.

By conducting these experiments far from the axial COM mode frequency $\omega_{z}$, we determine the measurement imprecision of this technique free from thermal and frequency noise of the COM mode.  For a crystal consisting of $N\sim88$ ions, a displacement amplitude of $49\,$pm is detected with a single measurement signal-to-noise ratio of $1$, which is an order-of-magnitude improvement over prior phase-incoherent experiments. This displacement is about $40$ times smaller than
\begin{equation} \label{ZPT}
    Z_{ZPT} = \frac{1}{\sqrt{N}}\sqrt{\frac{\hbar}{2m\omega_{z}}}\approx2\,\mathrm{nm}.
\end{equation}
With our repetition rate, this corresponds to a $8.4\,$pm$/\sqrt{\mathrm{Hz}}$ sensitivity.

The displacement sensitivity of this protocol implies a $12\,$yN$/\mathrm{ion}/\sqrt{\mathrm{Hz}}$ force sensitivity and a $77\,\mu$V$/$m$/\sqrt{\mathrm{Hz}}$ electric field sensitivity.  Rydberg atom sensors have similar electric field sensitivities~\cite{CoxRydberg}.  However, by sensing motion that is resonant with the axial COM mode, the sensitivity to weak forces and electric fields of this protocol could ideally be improved by the quality factor of the mode $Q\sim10^{6}$, indicating the possibility of electric field sensing below $\sim 1\,$nV/m in a few seconds.  

Electric field sensing below $\sim 1\,$nV/m enables searches for hidden-photon dark matter.  Ion traps typically operate with COM frequencies between 50 kHz and 5 MHz, providing a sensitivity to hidden photon masses from $2\times10^{-10}$ to $2\times10^{-8}$ eV. An electric field sensitivity of $0.35\,$nV/m corresponds to placing a limit on the kinetic mixing parameter of $\epsilon<10^{-13}$, which is more than an order-of-magnitude improvement over current cosmological limits over the $2\times10^{-10}$ to $2\times10^{-8}$ eV mass range~\cite{HornsDM, AriasDM, ChaudhuriDM}.  

We note that the above analysis does not consider shielding effects of the ion trap electrodes.  Specifically, these light mass hidden photons are characterized by long Compton wavelengths, and the interaction of the dark matter with the trap electrodes must also be considered.  For a dark matter search, the trap electrode and surrounding apparatus should be designed to minimize shielding effects.

The rest of the manuscript is structured as follows. In Sec.~\ref{ExpApp}, we describe the Penning trap setup, and the experimental protocol employed to measure COM motion of the crystal. Next, recent improvements that have enabled a relative phase stability between the driven motion and spin-dependent force of less than 5 degrees, including how this phase stability is measured at the ions are discussed in Sec.~\ref{PhaseStab}.  Section~\ref{DyDecoup} then focuses on the suppression of spin-decoherence through dynamical decoupling and the removal of an unwanted background interaction through a phase advance of the spin-dependent force.  In Sec.~\ref{Line}, theory and experimental measurements are shown for the line shapes of this sensing protocol.  Finally, in Sec.~\ref{Sens}, the main experimental result of this work, the order-of-magnitude improvement in the sensitivity to small displacements, is shown.  A brief summary and conclusion are included in Sec.~\ref{Con}.

\section{Experimental Apparatus}
\label{ExpApp}
\begin{figure}
\includegraphics[width=0.46\textwidth]{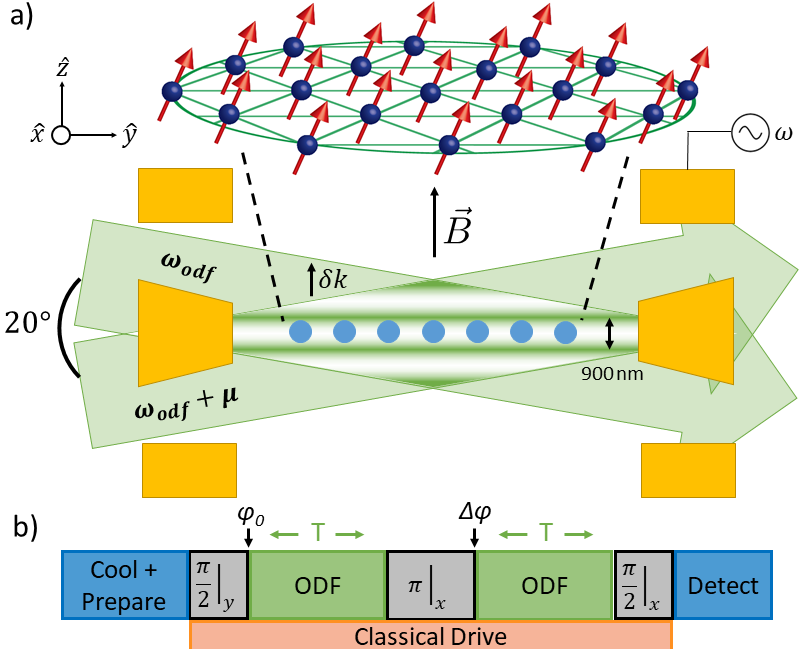}%
\caption{(a) An enlarged illustration of the 2D trapped-ion crystal, and simplified cross-sectional schematic of the Penning trap.  The ions (blue dots) have a typical ion-ion spacing of $15\,\mu$m resulting in an ion crystal with a radius of approximately $100\,\mu$m, and the electrodes (yellow) that generate the $\omega_{z}/2\pi=1.58\,$MHz axial confinement potential have a radius of $R_{w}=1\,$cm.  These electrodes are submerged in a $B=4.5\,$T magnetic field to confine the ions radially.  Shown in green are the two optical dipole force (ODF) beams intersecting at a $20^{\circ}$ angle to form the 1D traveling wave potential with a wavelength of $900\,$nm.  To produce a calibrated displacement of the ions, a rf potential is applied to one of the end cap electrodes.  (b) The motional amplitude is detected using a quantum lock-in~\cite{KotlerLockIn} spin-echo sequence.  The ions are initially cooled to $\bar{n}\sim1.6$ and prepared in the $\ket{\uparrow}_{N}$ spin state.  While the displacement is applied (orange block), a combination of spin-rotations (grey blocks) and application of the spin-dependent ODF (green blocks) induce spin-precession proportional to the displacement amplitude of the crystal, which is measured by recording the final fluorescence of the ions resulting from a pulse of the Doppler cooling laser.  The phase advance $\Delta \phi$ is chosen to chosen to maximize the spin-precession while canceling out a background signal.}
\label{TrapODF}
\end{figure}

These experiments are performed on 2D crystal arrays of about 100 $^{9}$Be$^{+}$ ions confined in a Penning trap.  Figure~\ref{TrapODF} (a) shows a simplified schematic of the trap and ion crystal array.  As discussed in prior work~\cite{BollingerApp,SawyerApp,BohnetApp}, radial confinement of the ions results from $E\times B$ induced rotation through the $B = 4.5\,\hat{z}$ Tesla magnetic field, and axial confinement is provided by a static quadratic potential along the z-axis formed by the stack of cylindrical electrodes.  The strength of this axially confining potential is quantified by the center-of-mass oscillation frequency $\omega_{z}/(2\pi)=1.58\,$MHz, which corresponds to the highest frequency axial mode~\cite{SawyerModes}.

In order to calibrate the displacement sensitivity of this technique, we oscillate the crystal axially at a frequency $\omega/(2\pi) \sim 400\,$kHz that is far from any mode by applying an rf sinusoidal potential to one of the end confining electrodes.  The displacement amplitude of the ions $Z_{c}$ is calibrated by measuring the shift in the equilibrium position of the crystal on a side-view imaging system when a large, static potential is applied.  A $1\,$V offset results in a $0.97(5)\,\mu$m displacement of the ions.  We estimate that the corrections for using this dc calibration to estimate $Z_{c}$ for an $\omega/(2\pi) \sim 400\,$kHz drive is less than $10\%$.

This axial displacement is detected using an experimental sequence similar to that shown in Fig.~\ref{TrapODF} (b).  The ions are initially cooled with a combination of Doppler and electromagnetically induced transparency (EIT) cooling.  Doppler cooling reduces the COM mode temperature to near the Doppler limit corresponding to an average motional quantum number of $\bar{n}=4.5$.  With the addition of EIT cooling, the COM mode can be cooled to $\bar{n}=0.3\pm0.2$, and all the other axial modes cooled to near their motional ground state~\cite{JordanEIT, AthreyaEIT}.  For this work, we intentionally weakened the EIT cooling to produce a long-term stable $\bar{n}\sim1.6$ COM mode temperature.  The addition of EIT cooling improves our sensitivity to small displacements by improving the Lamb-Dicke confinement, which increases the effective spin-dependent force.

After cooling, the spin state of the ions is initialized.  For our experiments, the two qubit states are the $\ket{\uparrow}\equiv|1/2,+1/2\rangle$ and $\ket{\downarrow}\equiv|1/2,-1/2\rangle$ valence electron spin projections of the $^{2}S_{1/2}$ electronic ground state.  The ions are initialized in the $\ket{\uparrow}$ state with a repump laser, and global spin rotations are implemented with a $\sim124\,$GHz microwave source.  The typical spin-echo sequence shown in Fig.~\ref{TrapODF} (b) consists of $\pi/2$ and $\pi$ pulses about the given Bloch sphere axes with a $\pi-$pulse duration $t_{\pi}$ of about 70~$\mu s$.  Additional $\pi-$pulses are included for longer experimental sequences to maintain a short free evolution duration $T$ in order to cancel noise from magnetic field fluctuations at frequencies below $T^{-1}$.  A final pulse of light on the Doppler cooling transition enables global readout of the qubit state.  In particular, the collected fluorescence is directly proportional to the fraction of ions in the $\ket{\uparrow}$ state, which will be referred to as the bright fraction $P_{\uparrow}$ throughout this paper.

To couple the spins and axial motion of the crystal, we implement a spin-dependent optical dipole force (ODF).  By overlapping two far-detuned $\sim313\,$nm beams at a $20^{\circ}$ angle at the ions as shown in Fig.~\ref{TrapODF} (a), a 1D optical lattice with an effective wavelength of $900\,$nm and a tunable beat frequency $\mu$ is formed.  This optical lattice couples the spin and motional degrees of freedom through the interaction
\begin{equation} \label{FullHam}
    \hat{H}_{\mathrm{ODF}}=U\sum_{i}\sin(\Delta k \hat{z}_{i}-\mu t-\phi_{0})\hat{\sigma}_{z},
\end{equation}
where $U$, $\Delta k$, and $\phi_{0}$ are the potential, wave vector, and phase of the optical lattice, respectively, and $\hat{z}_{i}$ is the axial position of ion $i$.  

Equation~\ref{FullHam} can be broken into two parts,
\begin{equation} \label{TwoHam}
\begin{aligned}
    \hat{H}_{\mathrm{ODF}}&=U\sum_{i}\sin(\Delta k \hat{z}_{i})\cos(\mu t+\phi_{0})\hat{\sigma}_{z} \\
    &-U\sum_{i}\cos(\Delta k \hat{z}_{i})\sin(\mu t+\phi_{0})\hat{\sigma}_{z}.
\end{aligned}
\end{equation}
The first term of Eq.~\ref{TwoHam} is first-order sensitive to the axial motion of the crystal, whereas the second term is a background interaction first-order independent of the $\hat{z}_{i}$.  This term is ignored in most treatments since the spin-precession resulting from this interaction is bounded by $(U/\hbar)/\mu$, which is typically small.  However, at low frequencies this background cannot be neglected.  We will show in Sec.~\ref{DyDecoup} that this background is decoupled from the spin-motion interaction by using an appropriate phase advance $\Delta\phi$ of the ODF in the spin-echo sequence.

With the background interaction removed using dynamical decoupling (see Sec.~\ref{DyDecoup}), we further simplify
this Hamiltonian by modelling the ion position as a driven COM displacement of the ion crystal with amplitude $Z_{c}$ at a frequency $\omega$ on top of the axial thermal motion of the ions, such that $\hat{z}_{i}\rightarrow \hat{z}_{i}+Z_{c}\cos(\omega t+\delta_{0})$.  Using $\Delta kZ_{c}\ll1$ and neglecting high frequency terms, Eq.~\ref{TwoHam} is simplified to 
\begin{equation} \label{SpinMoHam}
    \hat{H}_{\mathrm{ODF}}=F_{0}Z_{c}\cos[\Delta\mu t + \delta]\sum_{i}\frac{\sigma_{i}^{z}}{2},
\end{equation}
where $\Delta\mu = \omega-\mu$, $\delta\equiv\delta_{0}-\phi_{0}$ is the phase of the classical drive relative to the ODF phase, $F_{0}\equiv U \Delta k\, \mathrm{DWF}$ is the spin-dependent force, and the Debye-Waller factor $\mathrm{DWF}\equiv \mathrm{exp}[-\langle(\Delta k \hat{z}_{i})^{2}\rangle/2]$ reduces this force due to the departure from the Lamb-Dicke confinement regime.  With the modest EIT cooling used here, $\mathrm{DWF}\sim0.92$, and the spin-dependent force $F_{0}\sim88\,$yN.  This interaction produces a static shift in the spin transition frequency when $\Delta\mu=0$, and therefore gives rise to spin-precession of $\theta=\theta_{max}\cos(\delta)$, where $\theta_{max}=(F_{0}/\hbar)Z_{c}\tau$ and $\tau$ is the total duration in which spin-precession is accumulated.  It is through this spin-precession that small displacements of the ion crystal are measured.  

A spin-dependent force will drive spin-dependent motion.  We neglect this back-action throughout this manuscript.  This treatment is valid for the off-resonant sensing measurements discussed here because the differential displacements of the different spin states are always small compared to the ground state wave function size $Z_{ZPT}$~\cite{SawyerApp}.  For sensing motion that is resonant with the COM mode, the back action due to the driven spin-dependent motion cannot be neglected but can be avoided~\cite{HempelIon, ToscanoSub} when measuring the amplitude of a single quadrature of motion.

\section{Phase Stability}
\label{PhaseStab}
In prior implementations of this sensing protocol~\cite{GilmoreSensing}, the shot-to-shot phase of the ODF was incoherent with the axial motion of the crystal.  Due to this phase incoherence, an experimental sequence (final microwave pulse about the y-axis instead of x-axis) that was only second-order sensitive to displacements of the crystal was employed, and the ultimate sensitivity was reduced.  This section details the recent improvements in the stability of the phase $\delta$, which has enabled the more sensitive phase-coherent protocol, and opens new avenues of research including parametric amplification of the spin-motion interaction~\cite{BurdPara, GePara}.

Phase stability of the ODF at the ions requires stabilizing both the phase of the ODF beatnote and the axial position of the ions.  The wavelength of the ODF optical lattice is $900\,$nm, so to achieve a phase stability better than $10^{\circ}$ the equilibrium position of the ions must be maintained to within $25\,$nm.  For our trap, this requires limiting drifts in the approximately $2\,$kV confining potentials to less than $5\,$mV.  Also, the vibrational noise of the apparatus is reduced by floating the table.  To stabilize the ODF phase, the beam paths are enclosed to reduce interfermetric drifts between the two ODF beams, and the ODF beatnote is sensed and feedback stabilized before the beams enter the room temperature bore of the superconducting magnet ($\sim1\,$m from the ions).  

\begin{figure}
\includegraphics[width=0.46\textwidth]{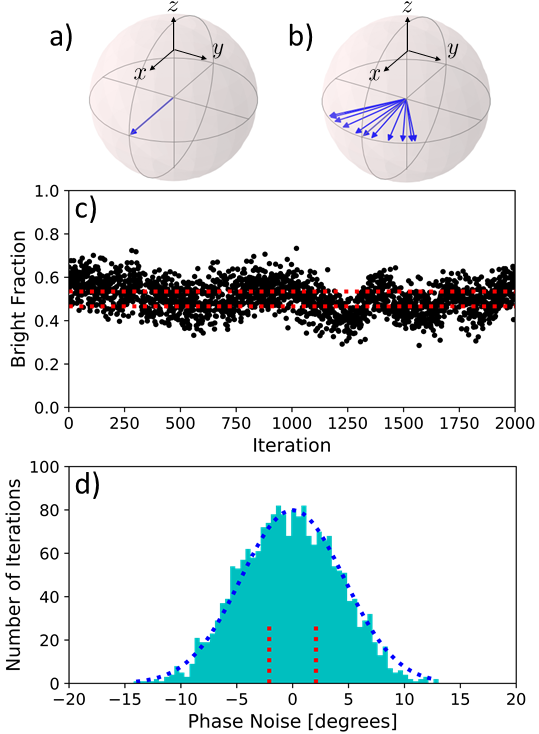}%
\caption{To measure the phase stability of the ODF at the ions, a spin-echo sequence is used in which the relative phase between the driven displacement and ODF is initially set to $\delta=\pi/2$.  With the ODF and driven displacement out-of-phase, no spin-precession occurs in either application of the ODF.  Therefore, in the absence of phase noise (a), the Bloch vector remains pointing along the x-axis.  However, in the presence of phase noise (b), spin-precession is accumulated, and the Bloch vector undergoes different amounts of spin precession from one trial of the experiment to the next.  (c) After the final $\pi/2$-pulse about the x-axis, the shot-to-shot variations in the spin-precession are rotated into noise in the measured bright fraction (black data) over that expected from spin projection noise (red dashed lines).  (d) Using a theory model, this increase in the noise in the bright fraction is interpreted as $5^{\circ}$ Gaussian phase noise between the driven motion and oscillating spin-dependent force (dashed blue curve).}
\label{PhaseStabFig}
\end{figure}

The phase stability of the ODF at the ions is measured using the spin-echo sequence shown in Fig.~\ref{TrapODF} (b).  These experiments are conducted on-resonance $\Delta\mu=0$ with the driven displacement and with a phase advance $\Delta\phi=\pi$, so that the spin-precession is accumulated in each arm.  To be first-order sensitive to the phase noise of the ODF, the initial relative phase is set to $\delta=\pi/2$.  Also, a large classical drive amplitude ($Z_{c}\sim5.5\,$nm) is used with $\theta_{max}\sim0.7\pi$ to improve the phase noise sensitivity.  

If from one realization of the experiment to the next the phase $\delta=\pi/2$ remains constant, then no spin-precession occurs in either of the ODF arms, and the Bloch vector remains along the x-axis for each trial as shown in Fig.~\ref{PhaseStabFig} (a).  Then, following the final $\pi/2$-pulse about the x-axis, the noise in the detected bright fraction is limited to spin projection noise.  However, if the phase of the ODF varies from one experimental trial to the next, then the resulting spin-precession causes dephasing of the Bloch vector for different iterations of the experiment as shown in Fig.~\ref{PhaseStabFig} (b), which results in increased noise in the bright fraction after the final $\pi/2$-pulse.

Figure~\ref{PhaseStabFig} (c) shows the resulting bright fraction of a $210 \pm 20$ ion cloud over $2000$ iterations of this phase stability experiment.  Each experiment lasts $9\,$ms, so this set of experiments analyzes the stability over a $18$ second interval.  The red dashed lines represent the standard deviation of spin projection noise.  Before improving the phase stability, the bright fraction wandered from full bright to full dark over the course of a few seconds.  After improving the laser phase stability and motional stability of the ions, the phase remains relatively constant over the $18$ second interval as shown by the data in Fig.~\ref{PhaseStabFig} (c) with some fast shot-to-shot noise that increases the standard deviation of the bright fraction by about a factor of two over projection noise.

From this increased noise in the bright fraction, we calculate the corresponding phase noise of the ODF.  The population of spins in the $\ket{\uparrow}$ state at the end of this experimental sequence is
\begin{equation} \label{Pup}
    P_{\uparrow}=\frac{1}{2}\left[1-e^{-\Gamma\tau}\sin(\theta)\right],
\end{equation}
where $\theta=\theta_{max}\cos(\delta)$ when $\Delta\mu=0$ and $\Gamma$ is the rate of spin-decoherence, which is predominantly due to off-resonant light scatter from the ODF beams.  We assume $\delta=\pi/2+\Delta\theta$, where the shot-to-shot phase noise $\Delta\theta$ is small.  Solving Eq.~\ref{Pup} for $\Delta\theta$, we find
\begin{equation} \label{PhNoise}
    \Delta\theta \approx \frac{e^{\Gamma\tau}\left(1-2P_{\uparrow}\right)}{\theta_{max}}.
\end{equation}
Figure~\ref{PhaseStabFig} (d) is a histogram of the phase noise for the data shown in Fig.~\ref{PhaseStabFig} (c).  The dashed vertical red lines correspond to the $2^{\circ}$ standard deviation from converting projection noise to phase noise, and the blue dashed curve is a Gaussian fit to the measured phase noise with a standard deviation of about $5$ degrees.  This measurement includes projection noise, and is therefore a conservative estimate of the phase stability.  With this phase stability, other sources of noise and background offsets are the limiting factor to these sensing experiments (see Sec.~\ref{Sens}).

\section{Dynamical Decoupling}
\label{DyDecoup}
\begin{figure}
\includegraphics[width=0.46\textwidth]{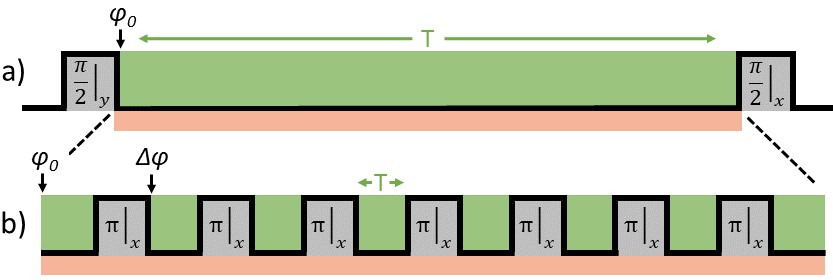}%
\caption{a) A sketch of the equivalent Ramsey sequence of this experimental protocol consisting of microwave pulses (grey boxes), and application of the ODF (green boxes) while the ion crystal is being driven by a classical drive (orange box).  For the given ODF power, a long free-evolution duration is required to obtain the optimum sensitivity to small displacements $(\tau=T)$.  b) To suppress magnetic field fluctuations, the free-precession period is divided into smaller sections by $\pi$-pulses about the x-axis with appropriate phase advances of the ODF $(\tau=8T)$.  The duration of these pulses are not drawn to scale.}
\label{Sequence}
\end{figure}

Dynamical decoupling is broadly used in quantum sensing and quantum computing to suppress decoherence from external sources of noise.  In the experiments reported here, the optimum sensitivity to small displacements requires a Ramsey sequence with a long free-precession duration (on the order of several milliseconds) as shown in Fig.~\ref{Sequence} (a), which makes this protocol susceptible to spin decoherence from magnetic field fluctuations.  To mitigate this effect, we employ multiple $\pi$-pulses about the x-axis as shown in Fig.~\ref{Sequence} (b) to shorten the free-precession time suppressing magnetic field noise for frequencies below $T^{-1}$.  With no phase advance of the ODF $\Delta\phi$ after these microwave $\pi$-pulses, this spin-echo protocol would also cancel the precession from the spin-motion coupling.  However, with the proper $\Delta\phi$, the desired spin-precession signal is accumulated in each arm.

In addition to suppressing spin-decoherence, the background interaction produced by the second term in Eq.~\ref{TwoHam} is removed in this protocol by applying the appropriate $\Delta\phi$.  For simplicity, we derive this cancellation for the simple spin-echo sequence shown in Fig.~\ref{TrapODF} (b), but this derivation holds for sequences with an odd number of $\pi$-pulses.  The total spin-precession accumulated from this background interaction is
\begin{equation} \label{Bck}
\begin{aligned}
    \theta_{bck}&=-\frac{U}{\hbar}\int_{t_{0}}^{t_{1}}\sin(\mu t+\phi_{0})dt\\
    &+\frac{U}{\hbar}\int_{t_{2}}^{t_{3}}\sin(\mu t+\phi_{0}-\Delta\phi)dt,
\end{aligned}
\end{equation}
where $t_{0}=t_{d}$, $t_{1}=T$, $t_{2}=T+t_{\pi}+t_{d}$, and $t_{3}=2T+t_{\pi}$ are the times at which the ODF lasers are turned off/on, and $t_{d}$ is a delay between setting the ODF phase and the start of the sequence.  In the phase-incoherent protocol, this delay led to a negligible contribution to the background, and was therefore neglected.  This phase-coherent protocol, however, is more sensitive, so $t_{d}$ cannot be ignored.  The resulting spin-precession is then
\begin{equation} \label{BckInt}
\begin{aligned}
    \theta_{bck}&=\frac{U}{\mu\hbar}\{\cos[\mu T+\phi_{0}]-\cos[\mu t_{d}+\phi_{0}]\\
    &+\cos[\mu (T+t_{\pi}+t_{d})+\phi_{0}-\Delta\phi]\\
    &-\cos[\mu (2T+t_{\pi})+\phi_{0}-\Delta\phi]\}.
\end{aligned}
\end{equation}
By applying a phase advance $\Delta\phi=\mu(T+t_{\pi})$, all of the terms in Eq.~\ref{BckInt} cancel, and therefore, this background interaction is removed for arbitrary $\mu$.  This analysis holds for sequences with an odd number of $\pi$-pulses, and a subsequent phase advance $\Delta\phi=m\mu(T+t_{\pi})$ after the $m^{th}$ $\pi$-pulse.

To maintain the spin-motion interaction of Eq.~\ref{TwoHam} under this phase advance, the classical drive must be applied at a particular frequency or the duration of the experiment must be tuned.  When 
\begin{equation} \label{ODFFreq}
    \frac{\mu}{2\pi} = \frac{2n + 1}{2(T + t_{\pi})}
\end{equation}
for some integer $n$, $\Delta\phi=\pi$ and the quantum lock-in phase advance of Ref.~\citenum{KotlerLockIn} is recovered. This phase advance coherently accumulates spin-precession from the first term of Eq.~\ref{TwoHam} when we apply the classical drive at 
\begin{equation} \label{DriveFreq}
    \frac{\omega}{2\pi} = \frac{2n + 1}{2(T + t_{\pi})},
\end{equation}
which for the work discussed here was approximately $400\,\mathrm{kHz}$.

\section{Line Shapes}
\label{Line}
When $\Delta\mu\ne0$, spin-precession from the spin-motion interaction of Eq.~\ref{SpinMoHam} still occurs, but the phase evolution throughout the sequence must be taken into account as well.  This results in a characteristic line shape (or response function) for the given experimental sequence. In this section, we theoretically and experimentally analyze the line shape from a constant amplitude driven displacement.  For simplicity, the derivation assumes the spin-echo sequence shown in Fig.~\ref{TrapODF} (b), but using trigonometric identities the phase factors reduce to the same analytical expression for spin-echo sequences with $m=1,3,7,15,...$ $\pi$-pulses.  We also neglect $t_{d}$ here, since $T\gg t_{d}$, so the effect on the signal is negligible.

The spin-precession accumulated in a general sequence is
\begin{equation} \label{SpinPre}
\begin{aligned}
    \theta(\mu)&=\frac{F_{0}Z_{c}}{\hbar}\int\cos\left(\Delta\mu t+\delta\right)dt \\
    &=\frac{F_{0}Z_{c}}{\hbar}\frac{2\sin\left(\frac{\Delta\mu}{2}T\right)}{\Delta\mu}\chi(\mu,\omega),
\end{aligned}
\end{equation}
where the phase evolution of each ODF arm is included in $\chi(\mu,\omega)=\sum_{j}\chi_{j}(\mu,\omega)$ and is determined by the particular sequence that is used.  For an application of the spin-motion coupling starting at $t_{i}$ and ending at $t_{f}$
\begin{equation} \label{ChiGen}
    \chi_{j}=\cos\left[\frac{\Delta\mu}{2}\left(t_{i}+t_{f}\right)+\delta\right],
\end{equation}
so for the two arms of the spin-echo sequence with the timings and phase advance discussed in Sec.~\ref{DyDecoup}
\begin{equation} \label{Chi1}
    \chi_{1}=\cos\left[\frac{\Delta\mu}{2}T+\delta\right],
\end{equation}
and
\begin{equation} \label{Chi2}
    \chi_{2}=\cos\left[\frac{\Delta\mu}{2}\left(3T+2t_{\pi}\right)+\mu\left(T+t_{\pi}\right)+\delta\right].
\end{equation}
Summing these terms and using the drive frequency of Eq.~\ref{DriveFreq}, we find
\begin{equation} \label{SpinPreLin}
    \theta(\mu)=\frac{F_{0}Z_{c}\tau}{\hbar}\mathrm{sinc}\left(\frac{\Delta\mu}{2}T\right)\cos\left(\frac{\Delta\mu}{2}T+\delta\right),
\end{equation}
where $\tau=(m+1)T$ is the total time the ODF is applied.  Again, this expression is dependent on the particular sequence used, but is valid for the spin-echo sequences with $m=1,3,7,15,...$ $\pi$-pulses discussed in this paper.

\begin{figure}
\includegraphics[width=0.46\textwidth]{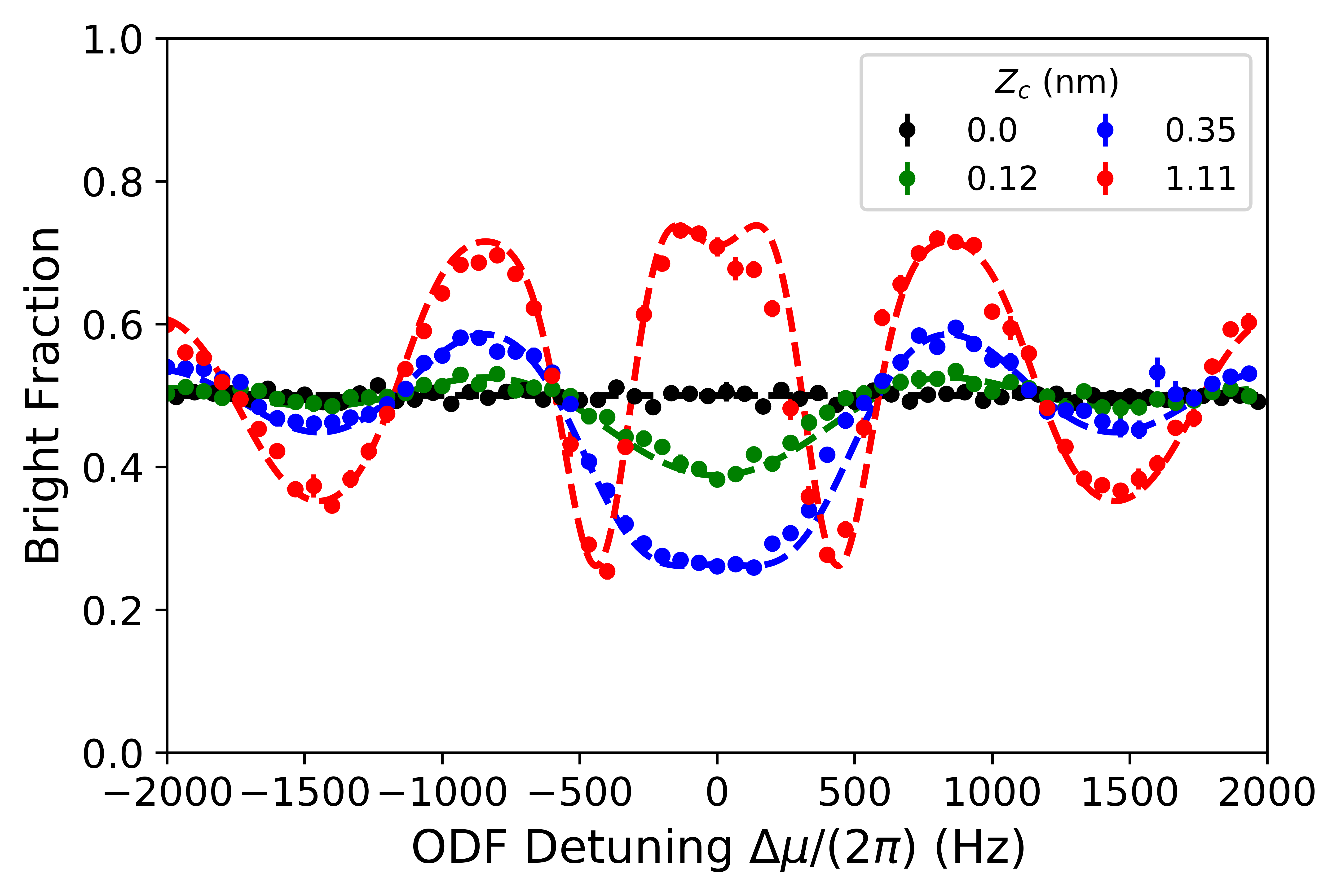}%
\caption{Measured bright fraction versus ODF detuning for various displacement amplitudes (symbols).  The error bars represent one standard deviation of uncertainty.  These line shapes are for a $m=7$ spin-echo sequence with an arm time $T=850\,\mu$s and the 2D crystal array consists of $N\sim69$ ions.  They are well described by theory (curves) given by Eqs.~\ref{Pup} and~\ref{SpinPreLin} with no adjustable parameters.}
\label{Lineshapes}
\end{figure}

Shown in Fig.~\ref{Lineshapes} are the measured line shapes for a spin-echo sequence with $m=7$ $\pi$-pulses.  When no drive is applied ($Z_{c}=0$), the bright fraction remains near 0.5 independent of the ODF frequency.  As the drive amplitude is increased, a signal emerges from the background.  On-resonance with $\delta=0$, the bright fraction is decreased for small drive amplitudes.  In contrast, for the largest drive amplitude ($Z_{c}=1.11\,$nm), the bright fraction is increased on-resonance since the induced spin-precession exceeds $\pi$.

The curves shown in Fig.~\ref{Lineshapes} are theory with no adjustable parameters.  Equation~\ref{Pup} is used to convert the calculated spin-precession accumulated for the spin-echo sequence (Eq.~\ref{SpinPreLin}) to a bright fraction.  The theory agrees well with the experiments for a range of drive amplitudes and ODF frequencies.

\section{Sensitivity}
\label{Sens}
\begin{figure}[t]
\includegraphics[width=0.46\textwidth]{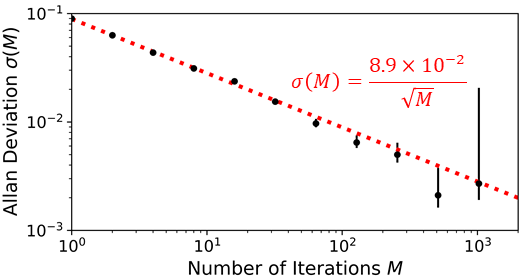}%
\caption{Allan deviation of the bright fraction measured over $\sim3000$ iterations of the experiment for the $Z_{c}=49\,$pm data set.  The fit (red dashed line) shows that the noise in the bright fraction is uncorrelated over this experimental interval, and therefore averages down as the square-root of the number of experiments.  Each iteration of the experiment (two $m=7$ spin-echo sequences) lasts about $30\,$ms.}
\label{Allan}
\end{figure}

Following a similar procedure as in Ref.~\citenum{GilmoreSensing}, we determine the ultimate amplitude sensing limit of this protocol by performing repeated pairs of $P_{\uparrow}$ measurements with the spin-dependent force applied at the same frequency as the classical drive.  Instead of using one of the two measurements to measure the background $(Z_{c}=0)$ as was done in Ref.~\citenum{GilmoreSensing}, the ability to control the relative phase between the classical drive and ODF allows us to advance the phase by $\pi$ between the first $P^{1}_{\uparrow}$ $(\delta=0)$ and second $P^{2}_{\uparrow}$ $(\delta=\pi)$ experiments.  This reverses the relative sign of the signal, and by taking the difference 
\begin{equation} \label{Signal}
    \langle P^{2}_{\uparrow}\rangle-\langle P^{1}_{\uparrow}\rangle=e^{-\Gamma\tau}\sin(\theta_{max}),
\end{equation}
we remove common offsets in the background and increase the size of the signal for this pair of experiments by a factor of two.  Equation~\ref{Signal} can be used to estimate $\theta_{max}$ and the displacement amplitude $Z_{c}$ through $\theta_{max}=F_0 Z_c \tau/\hbar$.  Figure~\ref{Allan} is a plot of the Allan deviation of the measured bright fraction $P^{2}_{\uparrow}-P^{1}_{\uparrow}$ for $3000$ iterations of this measurement.  The noise in these measurements averages down as the square root of the number of iterations M indicating good long-term stability of the experimental set-up. 

\begin{figure}
\includegraphics[width=0.46\textwidth]{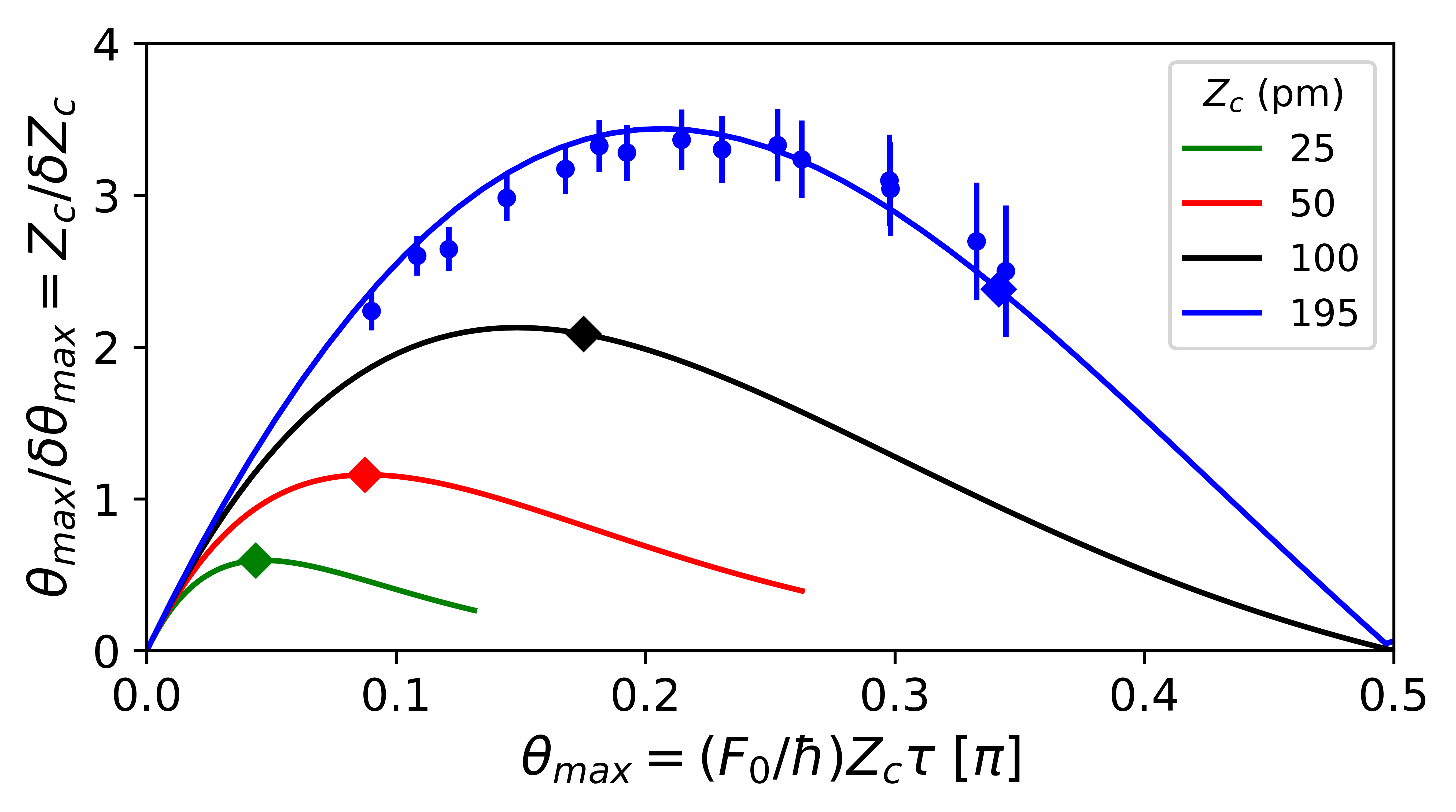}%
\caption{Amplitude sensitivity versus the angle of spin-precession for a range of displacement amplitudes $Z_{c}$.  Here $\tau=6.8\,$ms is fixed, and the $\theta_{max}$ is controlled by varying the strength of the ODF from $0$ to $3F_{0}^{\mathrm{max}}$.  The circles correspond to measurements with $Z_{c}=195\,$pm (error bars represent one standard deviation of uncertainty from repeated trials of the experiment), and the diamonds identify the location of maximum ODF power.  For $Z_{c}=195\,$pm, the optimum amplitude sensitivity under these conditions occurs for $\theta_{max}\sim0.2\pi$, which requires ODF strengths below the maximum.  As the displacement amplitude is decreased, the maximum amplitude sensitivity occurs at smaller spin-precession angles where the full ODF power is required.  The theory curves assume the experimentally observed $25\%$ increase in the observed noise over spin projection noise.}
\label{SNRTheta}
\end{figure}

The standard deviation $\delta \theta_{max}$ in estimating $\theta_{max}$ is determined by the standard deviation (ideally spin projection noise) $\sigma(P^{2}_{\uparrow}-P^{1}_{\uparrow})$ in the difference signal measurements through
\begin{equation} \label{Noise}
    \sigma(P^{2}_{\uparrow}-P^{1}_{\uparrow})=e^{-\Gamma\tau}\cos(\theta_{max})\delta\theta_{max}.
\end{equation}
Since 
\begin{equation} \label{AmpSen}
    \frac{Z_{c}}{\delta Z_{c}}=\frac{\theta_{max}}{\delta\theta_{max}}
\end{equation} 
the maximum sensitivity to small displacements occurs when $\theta_{max}/\delta\theta_{max}$ is maximized.  Using Eq.~\ref{AmpSen} and solving for $\theta_{max}$ and $\delta\theta_{max}$ from Eq.~\ref{Signal} and Eq.~\ref{Noise}, respectively, we can calculate the angle at which the optimal sensitivity is achieved.  Figure~\ref{SNRTheta} is a plot of the amplitude sensitivity versus the angle of spin-precession for a range of displacement amplitudes $Z_{c}$, where $\theta_{max}$ is controlled by varying $F_{0}$.  Note that increasing the strength of the ODF also increases the rate of spontaneous emission, which needs to be included when finding the angle.  Measurements and theory show that the optimum sensitivity to displacements occurs for $\theta_{max}\sim0.2\pi$ for the largest displacement amplitudes $Z_{c}\sim200\,$pm reported here.  For these large displacements, we lower the ODF power to remain at this optimum sensitivity.

Small displacements require higher ODF power.  Higher ODF power increases the impact of spontaneous emission, and $\theta_{max}/\delta\theta_{max}$ is maximized at small $\theta_{max}$ as shown in Fig.~\ref{SNRTheta} where small angle approximations to Eqs.~\ref{Signal} and \ref{Noise} are valid. For small $\theta_{max}$,
\begin{equation} \label{AmpSenSmall}
    \frac{\theta_{max}}{\delta\theta_{max}}\approx \frac{\langle P^{2}_{\uparrow}\rangle-\langle
    P^{1}_{\uparrow}\rangle}{\sigma(P^{2}_{\uparrow}-P^{1}_{\uparrow})}.
\end{equation} 
We define the experimentally determined signal-to-noise ratio SNR of a single pair of measurements as
\begin{equation} \label{SNR}
    \mathrm{SNR} \equiv\frac{\langle P^{2}_{\uparrow}\rangle-\langle
    P^{1}_{\uparrow}\rangle}{\sigma(P^{2}_{\uparrow}-P^{1}_{\uparrow})}.
\end{equation}
Therefore, for small amplitudes $Z_{c}$, the SNR provides a measure of the signal-to-noise ratio $Z_{c}/\delta Z_{c}$ for determining $Z_{c}$ in a single pair of measurements.

Assuming the noise is limited by spin projection noise such that $\delta\theta_{max}=e^{\Gamma\tau}/\sqrt{2N}$, we find the limiting amplitude sensitivity of this protocol to be
\begin{equation} \label{AmpLim}
    \left.\frac{Z_{c}}{\delta Z_{c}}\right\vert_{limiting}\approx \mathrm{DWF}(\Delta k Z_{c})\sqrt{2N}\frac{U\tau}{\hbar}e^{-\xi U\tau/\hbar},
\end{equation}
where $\xi=\Gamma/(U/\hbar)\sim1.14\times10^{-3}$ is the ratio of the spin-decoherence to the strength of the optical potential.  For a given number of ions, the amplitude sensitivity increases for longer applications of the ODF potential until spin-decoherence diminishes the contrast.  Equation~\ref{AmpLim} is maximized when $\Gamma\tau=1$, which for a typical $\Gamma\sim147\,$s$^{-1}$ sets $\tau\sim6.8\,$ms.  This motivates the duration of the protocol we implemented and corresponds to an ultimate amplitude sensitivity of
\begin{equation} \label{AmpUlt}
    \left.\frac{Z_{c}}{\delta Z_{c}}\right\vert_{ultimate}\approx \frac{F_{0}\tau}{\hbar e}\sqrt{2N}Z_{c}=\frac{Z_{c}}{36\,\mathrm{pm}},
\end{equation}
for the $N=88$ ions and $F_{0}=88\,$yN of these experiments. 

\begin{figure}
\includegraphics[width=0.46\textwidth]{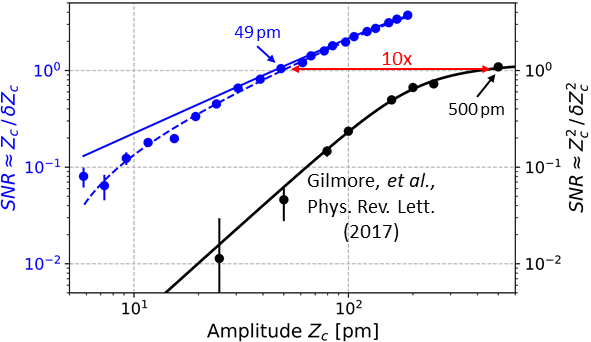}%
\caption{Amplitude sensing limits for a crystal of $N\sim88$ ions.  The black symbols and curve show the previous phase-incoherent measurements and projection noise limited theory of Ref.~\citenum{GilmoreSensing}, respectively. With this new phase-coherent scheme (blue symbols), a displacement amplitude of $49\,$pm is detected with a single measurement SNR of 1, which corresponds to an order-of-magnitude improvement in the sensitivity to small displacements.  At the smallest amplitudes, the SNR for the phase-incoherent scheme scales as $(Z_{c}/\delta Z_{c})^{2}$.  For the phase-coherent scheme, theory predicts first-order sensitivity to the displacement amplitude (solid blue curve).  However, at small amplitudes, the SNR measurements fall off faster than this prediction due to an offset in the background between the two $m=7$ spin-echo measurements (see text and Appendix~\ref{AppenA}).  We find good agreement between theory and experiment when this offset is included in the theory (blue dashed curve).  Both of these phase-coherent theory curves assume the measured $25\%$ increase in the background noise over projection noise.  The smallest detected amplitude with the $\sim3000$ experiments used here is $5.8\,$pm.  The error bars represent one standard deviation of uncertainty from repeated trails of the experiment.}
\label{SNRFig}
\end{figure}

Figure~\ref{SNRFig} shows the SNR for a single pair of measurements calculated from about $3000$ pairs of experiments for a wide range of displacement amplitudes.  The symbols and curve in black correspond to the previous phase-incoherent measurements and projection noise limited theory, respectively.  In those experiments, a $500\,$pm displacement amplitude was detected with a single measurement SNR of 1, and an amplitude of $50\,$pm was detected after averaging over the $3000$ pairs of experiments.  Due to the shot-to-shot phase noise inherent in this scheme, the SNR was limited to approximately 1 for amplitudes $Z_{c}\gtrsim500\,$pm.

With the phase-coherent protocol (blue data in Fig.~\ref{SNRFig}), a displacement amplitude of $49\,$pm is detected with a SNR of 1 with a single pair of measurements, which corresponds to an order-of-magnitude improvement in the amplitude sensitivity.  This amplitude is larger than that predicted by Eq.~\ref{AmpUlt} mainly due to additional noise in the bright fraction.  This excess noise most likely results from magnetic field fluctuations at frequencies above $T^{-1}$, and ideally a sequence with additional $\pi$-pulses (smaller $T$) would reduce this noise.  However, errors in the microwave pulses currently limits this protocol to $1.25\times$ spin-projection noise with a $m=7$ spin-echo sequence.

The solid blue curve of Fig.~\ref{SNRFig} is a full calculation of the SNR defined in Eq.~\ref{SNR} for the conditions of the experimental measurements and taking into account a $25\%$ increase in the experimental noise over spin projection noise.  The agreement is good for large amplitudes $(Z_{c}\gtrsim50\,$pm$)$.  The solid blue curve approaches the approximate result given by Eq.~\ref{AmpUlt} modified by the excess experimental noise for small angles $(Z_{c}\lesssim100\,$pm$)$.  At these smaller displacement amplitudes, the theory deviates from the experimental results.  This is due to an apparent $\sim2\%$ offset in the background between the first $P^{1}_{\uparrow}$ and second $P^{2}_{\uparrow}$ measurements.  This offset was determined by extrapolating a linear fit of the signal $\langle P^{2}_{\uparrow}\rangle-\langle P^{1}_{\uparrow}\rangle$ to zero drive amplitude (see Appendix~\ref{AppenA}).  We believe this offset is due to a small amount of cross talk between experimental control signals and the rf potential applied to the end cap electrode.  When the theory signal is reduced by this experimental offset, we have good agreement between theory and experiments for all displacement amplitudes (dashed blue curve).  Measurements for amplitudes $Z_{c} < 5.8\,$pm will require either a careful calibration of this offset or determining the source of the offset and getting rid of it entirely.

The slope of the SNR in Fig.~\ref{SNRFig} shows the benefit of the first-order amplitude scaling of this phase-coherent protocol over the second-order amplitude scaling of the prior phase-incoherent work~\cite{GilmoreSensing}.  If $Z_{c}$ is reduced by some factor $n$, the phase coherent scheme requires $n^{2}$ measurements to average down the noise.  In contrast, the phase incoherent scheme requires $n^{4}$ measurements.

Each iteration of this phase-coherent experiment consists of two $m=7$ spin-echo sequences, and lasts a total duration of about $30\,$ms.  Therefore, the displacement sensitivity of this technique is approximately $8.4\,$pm$/\sqrt{\mathrm{Hz}}$.  This implies force and electric field sensitivities of $12\,$yN$/\mathrm{ion}/\sqrt{\mathrm{Hz}}$ and $77\,\mu$V$/$m$/\sqrt{\mathrm{Hz}}$, respectively.

\section{Conclusion}
\label{Con}
In summary, we have shown experimental results in close agreement with theory for a new phase-coherent sensing protocol.  This technique relies on coupling axial motion to the internal spin degree of freedom of the ions through an oscillating spin-dependent force to produce spin-precession that is proportional to the amplitude of the motion.  With shot-to-shot noise in the phase between the driven displacement and ODF phase of less than $5^{\circ}$, a COM displacement of $49\,$pm was detected with a SNR of 1 in a single experimental determination.  This corresponds to an amplitude $40$ times smaller $Z_{ZPT}$, and a long measurement time sensitivity of $8.4\,$pm$/\sqrt{\mathrm{Hz}}$.

By performing these experiments far from the axial COM mode, we determine the measurement imprecision of this technique free from back-action and from thermal and frequency noise of this mode.  Moving on-resonance with the COM mode should improve the force and electric field sensitivity by several orders-of-magnitude, which would make this system one of the most sensitive quantum electric field sensors.  With electric field sensing below $\sim1\,$nV/m, searches for hidden-photon dark matter could be performed~\cite{HornsDM, AriasDM, ChaudhuriDM}.  Future work will explore the fundamental limits for on-resonance sensing including the effects of thermal and frequency fluctuations of the COM mode.

\begin{acknowledgments}
We thank David Hume, Robert Lewis-Swan, and Vivishek Sudhir for useful comments and discussions on this manuscript.  We acknowledge partial support from DOE Office of Science HEP QuantISED award.  M.A. was supported by an NRC fellowship funded by NIST. J.E.J. gratefully acknowledges the Leopoldina Research Fellowship, German National Academy of Sciences grant LPDS 2016-15, and the NIST-PREP program.
\end{acknowledgments}

\appendix
\section{Background Offset}
\label{AppenA}
\begin{figure}
\includegraphics[width=0.46\textwidth]{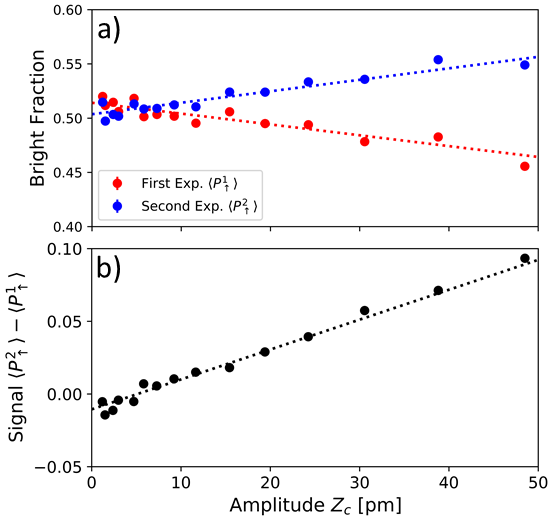}%
\caption{a) Symbols show the measured bright fraction for the first $P^{1}_{\uparrow}$ $(\delta=0)$ and second $P^{2}_{\uparrow}$ $(\delta=\pi)$ sensing experiments.  The scatter of these points around a linear fit (dashed lines) reflects the change in the background offset between successive experimental trials.  b) This scatter is reduced by using the difference in the two experiments as the experimental signal.  However, an offset in the background of approximately $2\%$ remains.  The error bars represent one standard deviation of uncertainty.}
\label{Offset}
\end{figure}

As discussed in Sec.~\ref{Sens}, an offset in the background between the first $P^{1}_{\uparrow}$ $(\delta=0)$ and second $P^{2}_{\uparrow}$ $(\delta=\pi)$ measurements impacts the current sensitivity of this protocol for determining small amplitudes.  Figure~\ref{Offset} (a) shows the measured bright fraction at small amplitudes for these two experiments.  The predicted linear dependence with $Z_{c}$ is observed.  However, linear fits (dashed curves) show an offset from expected background (bright fraction of $0.5$ at zero drive amplitude).

In Fig.~\ref{Offset} (b), we plot the difference between these two experiments, which is used as the experimental signal in Sec.~\ref{Sens}.  By taking the difference, we remove common drifts in the background as seen by the reduction in the scatter of the data away from the linear fit.  However, an approximate $2\%$ offset remains.  We believe this offset is due to a small amount of cross talk between experimental control signals and the rf potential applied to the end cap electrode.  This offset is rather robust since the data shown in Fig.~\ref{Offset} was taken over several hours on two different days.  Further investigation will be required to calibrate or reduce this offset.

\bibliographystyle{apsrev4-1}
\bibliography{bib}

\begin{thebibliography}{26}%
\makeatletter
\providecommand \@ifxundefined [1]{%
 \@ifx{#1\undefined}
}%
\providecommand \@ifnum [1]{%
 \ifnum #1\expandafter \@firstoftwo
 \else \expandafter \@secondoftwo
 \fi
}%
\providecommand \@ifx [1]{%
 \ifx #1\expandafter \@firstoftwo
 \else \expandafter \@secondoftwo
 \fi
}%
\providecommand \natexlab [1]{#1}%
\providecommand \enquote  [1]{``#1''}%
\providecommand \bibnamefont  [1]{#1}%
\providecommand \bibfnamefont [1]{#1}%
\providecommand \citenamefont [1]{#1}%
\providecommand \href@noop [0]{\@secondoftwo}%
\providecommand \href [0]{\begingroup \@sanitize@url \@href}%
\providecommand \@href[1]{\@@startlink{#1}\@@href}%
\providecommand \@@href[1]{\endgroup#1\@@endlink}%
\providecommand \@sanitize@url [0]{\catcode `\\12\catcode `\$12\catcode
  `\&12\catcode `\#12\catcode `\^12\catcode `\_12\catcode `\%12\relax}%
\providecommand \@@startlink[1]{}%
\providecommand \@@endlink[0]{}%
\providecommand \url  [0]{\begingroup\@sanitize@url \@url }%
\providecommand \@url [1]{\endgroup\@href {#1}{\urlprefix }}%
\providecommand \urlprefix  [0]{URL }%
\providecommand \Eprint [0]{\href }%
\providecommand \doibase [0]{http://dx.doi.org/}%
\providecommand \selectlanguage [0]{\@gobble}%
\providecommand \bibinfo  [0]{\@secondoftwo}%
\providecommand \bibfield  [0]{\@secondoftwo}%
\providecommand \translation [1]{[#1]}%
\providecommand \BibitemOpen [0]{}%
\providecommand \bibitemStop [0]{}%
\providecommand \bibitemNoStop [0]{.\EOS\space}%
\providecommand \EOS [0]{\spacefactor3000\relax}%
\providecommand \BibitemShut  [1]{\csname bibitem#1\endcsname}%
\let\auto@bib@innerbib\@empty
\bibitem [{\citenamefont {{ B. P. Abbott \textit{et al.} (LIGO Scientific
  Collaboration and Virgo Collaboration)}}(2016)}]{GravWave}%
  \BibitemOpen
  \bibfield  {author} {\bibinfo {author} {\bibnamefont {{ B. P. Abbott
  \textit{et al.} (LIGO Scientific Collaboration and Virgo Collaboration)}}},\
  }\href@noop {} {\bibfield  {journal} {\bibinfo  {journal} {Phys. Rev. Lett.}\
  }\textbf {\bibinfo {volume} {116}},\ \bibinfo {pages} {061102} (\bibinfo
  {year} {2016})}\BibitemShut {NoStop}%
\bibitem [{\citenamefont {Aspelmeyer}\ \emph {et~al.}(2014)\citenamefont
  {Aspelmeyer}, \citenamefont {Kippenberg},\ and\ \citenamefont
  {Marquardt}}]{AspelmeyerOpto}%
  \BibitemOpen
  \bibfield  {author} {\bibinfo {author} {\bibfnamefont {M.}~\bibnamefont
  {Aspelmeyer}}, \bibinfo {author} {\bibfnamefont {T.~J.}\ \bibnamefont
  {Kippenberg}}, \ and\ \bibinfo {author} {\bibfnamefont {F.}~\bibnamefont
  {Marquardt}},\ }\href {\doibase 10.1103/RevModPhys.86.1391} {\bibfield
  {journal} {\bibinfo  {journal} {Rev. Mod. Phys.}\ }\textbf {\bibinfo {volume}
  {86}},\ \bibinfo {pages} {1391} (\bibinfo {year} {2014})}\BibitemShut
  {NoStop}%
\bibitem [{\citenamefont {Wilson}\ \emph {et~al.}(2015)\citenamefont {Wilson},
  \citenamefont {Sudhir}, \citenamefont {Piro}, \citenamefont {Schilling},
  \citenamefont {Ghadimi},\ and\ \citenamefont {Kippenberg}}]{WilsonOpto}%
  \BibitemOpen
  \bibfield  {author} {\bibinfo {author} {\bibfnamefont {D.}~\bibnamefont
  {Wilson}}, \bibinfo {author} {\bibfnamefont {V.}~\bibnamefont {Sudhir}},
  \bibinfo {author} {\bibfnamefont {N.}~\bibnamefont {Piro}}, \bibinfo {author}
  {\bibfnamefont {R.}~\bibnamefont {Schilling}}, \bibinfo {author}
  {\bibfnamefont {A.}~\bibnamefont {Ghadimi}}, \ and\ \bibinfo {author}
  {\bibfnamefont {T.~J.}\ \bibnamefont {Kippenberg}},\ }\href@noop {}
  {\bibfield  {journal} {\bibinfo  {journal} {Nature}\ }\textbf {\bibinfo
  {volume} {524}},\ \bibinfo {pages} {325} (\bibinfo {year}
  {2015})}\BibitemShut {NoStop}%
\bibitem [{\citenamefont {Kampel}\ \emph {et~al.}(2017)\citenamefont {Kampel},
  \citenamefont {Peterson}, \citenamefont {Fischer}, \citenamefont {Yu},
  \citenamefont {Cicak}, \citenamefont {Simmonds}, \citenamefont {Lehnert},\
  and\ \citenamefont {Regal}}]{KampelOpto}%
  \BibitemOpen
  \bibfield  {author} {\bibinfo {author} {\bibfnamefont {N.~S.}\ \bibnamefont
  {Kampel}}, \bibinfo {author} {\bibfnamefont {R.~W.}\ \bibnamefont
  {Peterson}}, \bibinfo {author} {\bibfnamefont {R.}~\bibnamefont {Fischer}},
  \bibinfo {author} {\bibfnamefont {P.-L.}\ \bibnamefont {Yu}}, \bibinfo
  {author} {\bibfnamefont {K.}~\bibnamefont {Cicak}}, \bibinfo {author}
  {\bibfnamefont {R.~W.}\ \bibnamefont {Simmonds}}, \bibinfo {author}
  {\bibfnamefont {K.~W.}\ \bibnamefont {Lehnert}}, \ and\ \bibinfo {author}
  {\bibfnamefont {C.~A.}\ \bibnamefont {Regal}},\ }\href {\doibase
  10.1103/PhysRevX.7.021008} {\bibfield  {journal} {\bibinfo  {journal} {Phys.
  Rev. X}\ }\textbf {\bibinfo {volume} {7}},\ \bibinfo {pages} {021008}
  (\bibinfo {year} {2017})}\BibitemShut {NoStop}%
\bibitem [{\citenamefont {Schreppler}\ \emph {et~al.}(2014)\citenamefont
  {Schreppler}, \citenamefont {Spethmann}, \citenamefont {Brahms},
  \citenamefont {Botter}, \citenamefont {Barrios},\ and\ \citenamefont
  {Stamper-Kurn}}]{SchrepplerOpto}%
  \BibitemOpen
  \bibfield  {author} {\bibinfo {author} {\bibfnamefont {S.}~\bibnamefont
  {Schreppler}}, \bibinfo {author} {\bibfnamefont {N.}~\bibnamefont
  {Spethmann}}, \bibinfo {author} {\bibfnamefont {N.}~\bibnamefont {Brahms}},
  \bibinfo {author} {\bibfnamefont {T.}~\bibnamefont {Botter}}, \bibinfo
  {author} {\bibfnamefont {M.}~\bibnamefont {Barrios}}, \ and\ \bibinfo
  {author} {\bibfnamefont {D.~M.}\ \bibnamefont {Stamper-Kurn}},\ }\href
  {\doibase 10.1126/science.1249850} {\bibfield  {journal} {\bibinfo  {journal}
  {Science}\ }\textbf {\bibinfo {volume} {344}},\ \bibinfo {pages} {1486}
  (\bibinfo {year} {2014})}\BibitemShut {NoStop}%
\bibitem [{\citenamefont {Mason}\ \emph {et~al.}(2019)\citenamefont {Mason},
  \citenamefont {Chen}, \citenamefont {Rossi}, \citenamefont {Tsaturyan},\ and\
  \citenamefont {Schliesser}}]{MasonOpto}%
  \BibitemOpen
  \bibfield  {author} {\bibinfo {author} {\bibfnamefont {D.}~\bibnamefont
  {Mason}}, \bibinfo {author} {\bibfnamefont {J.}~\bibnamefont {Chen}},
  \bibinfo {author} {\bibfnamefont {M.}~\bibnamefont {Rossi}}, \bibinfo
  {author} {\bibfnamefont {Y.}~\bibnamefont {Tsaturyan}}, \ and\ \bibinfo
  {author} {\bibfnamefont {A.}~\bibnamefont {Schliesser}},\ }\href@noop {}
  {\bibfield  {journal} {\bibinfo  {journal} {Nature Physics}\ }\textbf
  {\bibinfo {volume} {15}},\ \bibinfo {pages} {745} (\bibinfo {year}
  {2019})}\BibitemShut {NoStop}%
\bibitem [{\citenamefont {Maiwald}\ \emph {et~al.}(2009)\citenamefont
  {Maiwald}, \citenamefont {Leibfried}, \citenamefont {Britton}, \citenamefont
  {Bergquist}, \citenamefont {Leuchs},\ and\ \citenamefont
  {Wineland}}]{MaiwaldIon}%
  \BibitemOpen
  \bibfield  {author} {\bibinfo {author} {\bibfnamefont {R.}~\bibnamefont
  {Maiwald}}, \bibinfo {author} {\bibfnamefont {D.}~\bibnamefont {Leibfried}},
  \bibinfo {author} {\bibfnamefont {J.}~\bibnamefont {Britton}}, \bibinfo
  {author} {\bibfnamefont {J.~C.}\ \bibnamefont {Bergquist}}, \bibinfo {author}
  {\bibfnamefont {G.}~\bibnamefont {Leuchs}}, \ and\ \bibinfo {author}
  {\bibfnamefont {D.~J.}\ \bibnamefont {Wineland}},\ }\href@noop {} {\bibfield
  {journal} {\bibinfo  {journal} {Nature Physics}\ }\textbf {\bibinfo {volume}
  {5}},\ \bibinfo {pages} {551} (\bibinfo {year} {2009})}\BibitemShut {NoStop}%
\bibitem [{\citenamefont {Hempel}\ \emph {et~al.}(2013)\citenamefont {Hempel},
  \citenamefont {Lanyon}, \citenamefont {Jurcevic}, \citenamefont {Gerritsma},
  \citenamefont {Blatt},\ and\ \citenamefont {Roos}}]{HempelIon}%
  \BibitemOpen
  \bibfield  {author} {\bibinfo {author} {\bibfnamefont {C.}~\bibnamefont
  {Hempel}}, \bibinfo {author} {\bibfnamefont {B.}~\bibnamefont {Lanyon}},
  \bibinfo {author} {\bibfnamefont {P.}~\bibnamefont {Jurcevic}}, \bibinfo
  {author} {\bibfnamefont {R.}~\bibnamefont {Gerritsma}}, \bibinfo {author}
  {\bibfnamefont {R.}~\bibnamefont {Blatt}}, \ and\ \bibinfo {author}
  {\bibfnamefont {C.}~\bibnamefont {Roos}},\ }\href@noop {} {\bibfield
  {journal} {\bibinfo  {journal} {Nature Photonics}\ }\textbf {\bibinfo
  {volume} {7}},\ \bibinfo {pages} {630} (\bibinfo {year} {2013})}\BibitemShut
  {NoStop}%
\bibitem [{\citenamefont {Shaniv}\ and\ \citenamefont
  {Ozeri}(2017)}]{ShanivIon}%
  \BibitemOpen
  \bibfield  {author} {\bibinfo {author} {\bibfnamefont {R.}~\bibnamefont
  {Shaniv}}\ and\ \bibinfo {author} {\bibfnamefont {R.}~\bibnamefont {Ozeri}},\
  }\href@noop {} {\bibfield  {journal} {\bibinfo  {journal} {Nature
  communications}\ }\textbf {\bibinfo {volume} {8}},\ \bibinfo {pages} {1}
  (\bibinfo {year} {2017})}\BibitemShut {NoStop}%
\bibitem [{\citenamefont {Burd}\ \emph {et~al.}(2019)\citenamefont {Burd},
  \citenamefont {Srinivas}, \citenamefont {Bollinger}, \citenamefont {Wilson},
  \citenamefont {Wineland}, \citenamefont {Leibfried}, \citenamefont
  {Slichter},\ and\ \citenamefont {Allcock}}]{BurdPara}%
  \BibitemOpen
  \bibfield  {author} {\bibinfo {author} {\bibfnamefont {S.~C.}\ \bibnamefont
  {Burd}}, \bibinfo {author} {\bibfnamefont {R.}~\bibnamefont {Srinivas}},
  \bibinfo {author} {\bibfnamefont {J.~J.}\ \bibnamefont {Bollinger}}, \bibinfo
  {author} {\bibfnamefont {A.~C.}\ \bibnamefont {Wilson}}, \bibinfo {author}
  {\bibfnamefont {D.~J.}\ \bibnamefont {Wineland}}, \bibinfo {author}
  {\bibfnamefont {D.}~\bibnamefont {Leibfried}}, \bibinfo {author}
  {\bibfnamefont {D.~H.}\ \bibnamefont {Slichter}}, \ and\ \bibinfo {author}
  {\bibfnamefont {D.~T.~C.}\ \bibnamefont {Allcock}},\ }\href {\doibase
  10.1126/science.aaw2884} {\bibfield  {journal} {\bibinfo  {journal}
  {Science}\ }\textbf {\bibinfo {volume} {364}},\ \bibinfo {pages} {1163}
  (\bibinfo {year} {2019})}\BibitemShut {NoStop}%
\bibitem [{\citenamefont {Gilmore}\ \emph {et~al.}(2017)\citenamefont
  {Gilmore}, \citenamefont {Bohnet}, \citenamefont {Sawyer}, \citenamefont
  {Britton},\ and\ \citenamefont {Bollinger}}]{GilmoreSensing}%
  \BibitemOpen
  \bibfield  {author} {\bibinfo {author} {\bibfnamefont {K.~A.}\ \bibnamefont
  {Gilmore}}, \bibinfo {author} {\bibfnamefont {J.~G.}\ \bibnamefont {Bohnet}},
  \bibinfo {author} {\bibfnamefont {B.~C.}\ \bibnamefont {Sawyer}}, \bibinfo
  {author} {\bibfnamefont {J.~W.}\ \bibnamefont {Britton}}, \ and\ \bibinfo
  {author} {\bibfnamefont {J.~J.}\ \bibnamefont {Bollinger}},\ }\href {\doibase
  10.1103/PhysRevLett.118.263602} {\bibfield  {journal} {\bibinfo  {journal}
  {Phys. Rev. Lett.}\ }\textbf {\bibinfo {volume} {118}},\ \bibinfo {pages}
  {263602} (\bibinfo {year} {2017})}\BibitemShut {NoStop}%
\bibitem [{\citenamefont {Wolf}\ \emph {et~al.}(2019)\citenamefont {Wolf},
  \citenamefont {Shi}, \citenamefont {Heip}, \citenamefont {Gessner},
  \citenamefont {Pezz{\`e}}, \citenamefont {Smerzi}, \citenamefont {Schulte},
  \citenamefont {Hammerer},\ and\ \citenamefont {Schmidt}}]{WolfFock}%
  \BibitemOpen
  \bibfield  {author} {\bibinfo {author} {\bibfnamefont {F.}~\bibnamefont
  {Wolf}}, \bibinfo {author} {\bibfnamefont {C.}~\bibnamefont {Shi}}, \bibinfo
  {author} {\bibfnamefont {J.~C.}\ \bibnamefont {Heip}}, \bibinfo {author}
  {\bibfnamefont {M.}~\bibnamefont {Gessner}}, \bibinfo {author} {\bibfnamefont
  {L.}~\bibnamefont {Pezz{\`e}}}, \bibinfo {author} {\bibfnamefont
  {A.}~\bibnamefont {Smerzi}}, \bibinfo {author} {\bibfnamefont
  {M.}~\bibnamefont {Schulte}}, \bibinfo {author} {\bibfnamefont
  {K.}~\bibnamefont {Hammerer}}, \ and\ \bibinfo {author} {\bibfnamefont
  {P.~O.}\ \bibnamefont {Schmidt}},\ }\href@noop {} {\bibfield  {journal}
  {\bibinfo  {journal} {Nature communications}\ }\textbf {\bibinfo {volume}
  {10}},\ \bibinfo {pages} {1} (\bibinfo {year} {2019})}\BibitemShut {NoStop}%
\bibitem [{\citenamefont {Biercuk}\ \emph {et~al.}(2010)\citenamefont
  {Biercuk}, \citenamefont {Uys}, \citenamefont {Britton}, \citenamefont
  {VanDevender},\ and\ \citenamefont {Bollinger}}]{BiercukMultIon}%
  \BibitemOpen
  \bibfield  {author} {\bibinfo {author} {\bibfnamefont {M.~J.}\ \bibnamefont
  {Biercuk}}, \bibinfo {author} {\bibfnamefont {H.}~\bibnamefont {Uys}},
  \bibinfo {author} {\bibfnamefont {J.~W.}\ \bibnamefont {Britton}}, \bibinfo
  {author} {\bibfnamefont {A.~P.}\ \bibnamefont {VanDevender}}, \ and\ \bibinfo
  {author} {\bibfnamefont {J.~J.}\ \bibnamefont {Bollinger}},\ }\href@noop {}
  {\bibfield  {journal} {\bibinfo  {journal} {Nature nanotechnology}\ }\textbf
  {\bibinfo {volume} {5}},\ \bibinfo {pages} {646} (\bibinfo {year}
  {2010})}\BibitemShut {NoStop}%
\bibitem [{\citenamefont {Ge}\ \emph {et~al.}(2019)\citenamefont {Ge},
  \citenamefont {Sawyer}, \citenamefont {Britton}, \citenamefont {Jacobs},
  \citenamefont {Bollinger},\ and\ \citenamefont {Foss-Feig}}]{GePara}%
  \BibitemOpen
  \bibfield  {author} {\bibinfo {author} {\bibfnamefont {W.}~\bibnamefont
  {Ge}}, \bibinfo {author} {\bibfnamefont {B.~C.}\ \bibnamefont {Sawyer}},
  \bibinfo {author} {\bibfnamefont {J.~W.}\ \bibnamefont {Britton}}, \bibinfo
  {author} {\bibfnamefont {K.}~\bibnamefont {Jacobs}}, \bibinfo {author}
  {\bibfnamefont {J.~J.}\ \bibnamefont {Bollinger}}, \ and\ \bibinfo {author}
  {\bibfnamefont {M.}~\bibnamefont {Foss-Feig}},\ }\href {\doibase
  10.1103/PhysRevLett.122.030501} {\bibfield  {journal} {\bibinfo  {journal}
  {Phys. Rev. Lett.}\ }\textbf {\bibinfo {volume} {122}},\ \bibinfo {pages}
  {030501} (\bibinfo {year} {2019})}\BibitemShut {NoStop}%
\bibitem [{\citenamefont {Meyer}\ \emph {et~al.}(2020)\citenamefont {Meyer},
  \citenamefont {Castillo}, \citenamefont {Cox},\ and\ \citenamefont
  {Kunz}}]{CoxRydberg}%
  \BibitemOpen
  \bibfield  {author} {\bibinfo {author} {\bibfnamefont {D.~H.}\ \bibnamefont
  {Meyer}}, \bibinfo {author} {\bibfnamefont {Z.~A.}\ \bibnamefont {Castillo}},
  \bibinfo {author} {\bibfnamefont {K.~C.}\ \bibnamefont {Cox}}, \ and\
  \bibinfo {author} {\bibfnamefont {P.~D.}\ \bibnamefont {Kunz}},\ }\href@noop
  {} {\bibfield  {journal} {\bibinfo  {journal} {Journal of Physics B: Atomic,
  Molecular and Optical Physics}\ }\textbf {\bibinfo {volume} {53}},\ \bibinfo
  {pages} {034001} (\bibinfo {year} {2020})}\BibitemShut {NoStop}%
\bibitem [{\citenamefont {Horns}\ \emph {et~al.}(2013)\citenamefont {Horns},
  \citenamefont {Jaeckel}, \citenamefont {Lindner}, \citenamefont {Lobanov},
  \citenamefont {Redondo},\ and\ \citenamefont {Ringwald}}]{HornsDM}%
  \BibitemOpen
  \bibfield  {author} {\bibinfo {author} {\bibfnamefont {D.}~\bibnamefont
  {Horns}}, \bibinfo {author} {\bibfnamefont {J.}~\bibnamefont {Jaeckel}},
  \bibinfo {author} {\bibfnamefont {A.}~\bibnamefont {Lindner}}, \bibinfo
  {author} {\bibfnamefont {A.}~\bibnamefont {Lobanov}}, \bibinfo {author}
  {\bibfnamefont {J.}~\bibnamefont {Redondo}}, \ and\ \bibinfo {author}
  {\bibfnamefont {A.}~\bibnamefont {Ringwald}},\ }\href {\doibase
  10.1088/1475-7516/2013/04/016} {\bibfield  {journal} {\bibinfo  {journal}
  {Journal of Cosmology and Astroparticle Physics}\ }\textbf {\bibinfo {volume}
  {2013}},\ \bibinfo {pages} {016} (\bibinfo {year} {2013})}\BibitemShut
  {NoStop}%
\bibitem [{\citenamefont {Arias}\ \emph {et~al.}(2012)\citenamefont {Arias},
  \citenamefont {Cadamuro}, \citenamefont {Goodsell}, \citenamefont {Jaeckel},
  \citenamefont {Redondo},\ and\ \citenamefont {Ringwald}}]{AriasDM}%
  \BibitemOpen
  \bibfield  {author} {\bibinfo {author} {\bibfnamefont {P.}~\bibnamefont
  {Arias}}, \bibinfo {author} {\bibfnamefont {D.}~\bibnamefont {Cadamuro}},
  \bibinfo {author} {\bibfnamefont {M.}~\bibnamefont {Goodsell}}, \bibinfo
  {author} {\bibfnamefont {J.}~\bibnamefont {Jaeckel}}, \bibinfo {author}
  {\bibfnamefont {J.}~\bibnamefont {Redondo}}, \ and\ \bibinfo {author}
  {\bibfnamefont {A.}~\bibnamefont {Ringwald}},\ }\href {\doibase
  10.1088/1475-7516/2012/06/013} {\bibfield  {journal} {\bibinfo  {journal}
  {Journal of Cosmology and Astroparticle Physics}\ }\textbf {\bibinfo {volume}
  {2012}},\ \bibinfo {pages} {013} (\bibinfo {year} {2012})}\BibitemShut
  {NoStop}%
\bibitem [{\citenamefont {Chaudhuri}\ \emph {et~al.}(2015)\citenamefont
  {Chaudhuri}, \citenamefont {Graham}, \citenamefont {Irwin}, \citenamefont
  {Mardon}, \citenamefont {Rajendran},\ and\ \citenamefont
  {Zhao}}]{ChaudhuriDM}%
  \BibitemOpen
  \bibfield  {author} {\bibinfo {author} {\bibfnamefont {S.}~\bibnamefont
  {Chaudhuri}}, \bibinfo {author} {\bibfnamefont {P.~W.}\ \bibnamefont
  {Graham}}, \bibinfo {author} {\bibfnamefont {K.}~\bibnamefont {Irwin}},
  \bibinfo {author} {\bibfnamefont {J.}~\bibnamefont {Mardon}}, \bibinfo
  {author} {\bibfnamefont {S.}~\bibnamefont {Rajendran}}, \ and\ \bibinfo
  {author} {\bibfnamefont {Y.}~\bibnamefont {Zhao}},\ }\href {\doibase
  10.1103/PhysRevD.92.075012} {\bibfield  {journal} {\bibinfo  {journal} {Phys.
  Rev. D}\ }\textbf {\bibinfo {volume} {92}},\ \bibinfo {pages} {075012}
  (\bibinfo {year} {2015})}\BibitemShut {NoStop}%
\bibitem [{\citenamefont {Bollinger}\ \emph {et~al.}(2013)\citenamefont
  {Bollinger}, \citenamefont {Britton},\ and\ \citenamefont
  {Sawyer}}]{BollingerApp}%
  \BibitemOpen
  \bibfield  {author} {\bibinfo {author} {\bibfnamefont {J.~J.}\ \bibnamefont
  {Bollinger}}, \bibinfo {author} {\bibfnamefont {J.~W.}\ \bibnamefont
  {Britton}}, \ and\ \bibinfo {author} {\bibfnamefont {B.~C.}\ \bibnamefont
  {Sawyer}},\ }\href {\doibase 10.1063/1.4796076} {\bibfield  {journal}
  {\bibinfo  {journal} {AIP Conference Proceedings}\ }\textbf {\bibinfo
  {volume} {1521}},\ \bibinfo {pages} {200} (\bibinfo {year}
  {2013})}\BibitemShut {NoStop}%
\bibitem [{\citenamefont {Sawyer}\ \emph {et~al.}(2014)\citenamefont {Sawyer},
  \citenamefont {Britton},\ and\ \citenamefont {Bollinger}}]{SawyerApp}%
  \BibitemOpen
  \bibfield  {author} {\bibinfo {author} {\bibfnamefont {B.~C.}\ \bibnamefont
  {Sawyer}}, \bibinfo {author} {\bibfnamefont {J.~W.}\ \bibnamefont {Britton}},
  \ and\ \bibinfo {author} {\bibfnamefont {J.~J.}\ \bibnamefont {Bollinger}},\
  }\href {\doibase 10.1103/PhysRevA.89.033408} {\bibfield  {journal} {\bibinfo
  {journal} {Phys. Rev. A}\ }\textbf {\bibinfo {volume} {89}},\ \bibinfo
  {pages} {033408} (\bibinfo {year} {2014})}\BibitemShut {NoStop}%
\bibitem [{\citenamefont {Bohnet}\ \emph {et~al.}(2016)\citenamefont {Bohnet},
  \citenamefont {Sawyer}, \citenamefont {Britton}, \citenamefont {Wall},
  \citenamefont {Rey}, \citenamefont {Foss-Feig},\ and\ \citenamefont
  {Bollinger}}]{BohnetApp}%
  \BibitemOpen
  \bibfield  {author} {\bibinfo {author} {\bibfnamefont {J.~G.}\ \bibnamefont
  {Bohnet}}, \bibinfo {author} {\bibfnamefont {B.~C.}\ \bibnamefont {Sawyer}},
  \bibinfo {author} {\bibfnamefont {J.~W.}\ \bibnamefont {Britton}}, \bibinfo
  {author} {\bibfnamefont {M.~L.}\ \bibnamefont {Wall}}, \bibinfo {author}
  {\bibfnamefont {A.~M.}\ \bibnamefont {Rey}}, \bibinfo {author} {\bibfnamefont
  {M.}~\bibnamefont {Foss-Feig}}, \ and\ \bibinfo {author} {\bibfnamefont
  {J.~J.}\ \bibnamefont {Bollinger}},\ }\href {\doibase
  10.1126/science.aad9958} {\bibfield  {journal} {\bibinfo  {journal}
  {Science}\ }\textbf {\bibinfo {volume} {352}},\ \bibinfo {pages} {1297}
  (\bibinfo {year} {2016})}\BibitemShut {NoStop}%
\bibitem [{\citenamefont {Sawyer}\ \emph {et~al.}(2012)\citenamefont {Sawyer},
  \citenamefont {Britton}, \citenamefont {Keith}, \citenamefont {Wang},
  \citenamefont {Freericks}, \citenamefont {Uys}, \citenamefont {Biercuk},\
  and\ \citenamefont {Bollinger}}]{SawyerModes}%
  \BibitemOpen
  \bibfield  {author} {\bibinfo {author} {\bibfnamefont {B.~C.}\ \bibnamefont
  {Sawyer}}, \bibinfo {author} {\bibfnamefont {J.~W.}\ \bibnamefont {Britton}},
  \bibinfo {author} {\bibfnamefont {A.~C.}\ \bibnamefont {Keith}}, \bibinfo
  {author} {\bibfnamefont {C.-C.~J.}\ \bibnamefont {Wang}}, \bibinfo {author}
  {\bibfnamefont {J.~K.}\ \bibnamefont {Freericks}}, \bibinfo {author}
  {\bibfnamefont {H.}~\bibnamefont {Uys}}, \bibinfo {author} {\bibfnamefont
  {M.~J.}\ \bibnamefont {Biercuk}}, \ and\ \bibinfo {author} {\bibfnamefont
  {J.~J.}\ \bibnamefont {Bollinger}},\ }\href {\doibase
  10.1103/PhysRevLett.108.213003} {\bibfield  {journal} {\bibinfo  {journal}
  {Phys. Rev. Lett.}\ }\textbf {\bibinfo {volume} {108}},\ \bibinfo {pages}
  {213003} (\bibinfo {year} {2012})}\BibitemShut {NoStop}%
\bibitem [{\citenamefont {Kotler}\ \emph {et~al.}(2011)\citenamefont {Kotler},
  \citenamefont {Akerman}, \citenamefont {Glickman}, \citenamefont {Keselman},\
  and\ \citenamefont {Ozeri}}]{KotlerLockIn}%
  \BibitemOpen
  \bibfield  {author} {\bibinfo {author} {\bibfnamefont {S.}~\bibnamefont
  {Kotler}}, \bibinfo {author} {\bibfnamefont {N.}~\bibnamefont {Akerman}},
  \bibinfo {author} {\bibfnamefont {Y.}~\bibnamefont {Glickman}}, \bibinfo
  {author} {\bibfnamefont {A.}~\bibnamefont {Keselman}}, \ and\ \bibinfo
  {author} {\bibfnamefont {R.}~\bibnamefont {Ozeri}},\ }\href@noop {}
  {\bibfield  {journal} {\bibinfo  {journal} {Nature}\ }\textbf {\bibinfo
  {volume} {473}},\ \bibinfo {pages} {61} (\bibinfo {year} {2011})}\BibitemShut
  {NoStop}%
\bibitem [{\citenamefont {Jordan}\ \emph {et~al.}(2019)\citenamefont {Jordan},
  \citenamefont {Gilmore}, \citenamefont {Shankar}, \citenamefont
  {Safavi-Naini}, \citenamefont {Bohnet}, \citenamefont {Holland},\ and\
  \citenamefont {Bollinger}}]{JordanEIT}%
  \BibitemOpen
  \bibfield  {author} {\bibinfo {author} {\bibfnamefont {E.}~\bibnamefont
  {Jordan}}, \bibinfo {author} {\bibfnamefont {K.~A.}\ \bibnamefont {Gilmore}},
  \bibinfo {author} {\bibfnamefont {A.}~\bibnamefont {Shankar}}, \bibinfo
  {author} {\bibfnamefont {A.}~\bibnamefont {Safavi-Naini}}, \bibinfo {author}
  {\bibfnamefont {J.~G.}\ \bibnamefont {Bohnet}}, \bibinfo {author}
  {\bibfnamefont {M.~J.}\ \bibnamefont {Holland}}, \ and\ \bibinfo {author}
  {\bibfnamefont {J.~J.}\ \bibnamefont {Bollinger}},\ }\href {\doibase
  10.1103/PhysRevLett.122.053603} {\bibfield  {journal} {\bibinfo  {journal}
  {Phys. Rev. Lett.}\ }\textbf {\bibinfo {volume} {122}},\ \bibinfo {pages}
  {053603} (\bibinfo {year} {2019})}\BibitemShut {NoStop}%
\bibitem [{\citenamefont {Shankar}\ \emph {et~al.}(2019)\citenamefont
  {Shankar}, \citenamefont {Jordan}, \citenamefont {Gilmore}, \citenamefont
  {Safavi-Naini}, \citenamefont {Bollinger},\ and\ \citenamefont
  {Holland}}]{AthreyaEIT}%
  \BibitemOpen
  \bibfield  {author} {\bibinfo {author} {\bibfnamefont {A.}~\bibnamefont
  {Shankar}}, \bibinfo {author} {\bibfnamefont {E.}~\bibnamefont {Jordan}},
  \bibinfo {author} {\bibfnamefont {K.~A.}\ \bibnamefont {Gilmore}}, \bibinfo
  {author} {\bibfnamefont {A.}~\bibnamefont {Safavi-Naini}}, \bibinfo {author}
  {\bibfnamefont {J.~J.}\ \bibnamefont {Bollinger}}, \ and\ \bibinfo {author}
  {\bibfnamefont {M.~J.}\ \bibnamefont {Holland}},\ }\href {\doibase
  10.1103/PhysRevA.99.023409} {\bibfield  {journal} {\bibinfo  {journal} {Phys.
  Rev. A}\ }\textbf {\bibinfo {volume} {99}},\ \bibinfo {pages} {023409}
  (\bibinfo {year} {2019})}\BibitemShut {NoStop}%
\bibitem [{\citenamefont {Toscano}\ \emph {et~al.}(2006)\citenamefont
  {Toscano}, \citenamefont {Dalvit}, \citenamefont {Davidovich},\ and\
  \citenamefont {Zurek}}]{ToscanoSub}%
  \BibitemOpen
  \bibfield  {author} {\bibinfo {author} {\bibfnamefont {F.}~\bibnamefont
  {Toscano}}, \bibinfo {author} {\bibfnamefont {D.~A.~R.}\ \bibnamefont
  {Dalvit}}, \bibinfo {author} {\bibfnamefont {L.}~\bibnamefont {Davidovich}},
  \ and\ \bibinfo {author} {\bibfnamefont {W.~H.}\ \bibnamefont {Zurek}},\
  }\href {\doibase 10.1103/PhysRevA.73.023803} {\bibfield  {journal} {\bibinfo
  {journal} {Phys. Rev. A}\ }\textbf {\bibinfo {volume} {73}},\ \bibinfo
  {pages} {023803} (\bibinfo {year} {2006})}\BibitemShut {NoStop}%
\end{thebibliography}%

\end{document}